\documentstyle[prb,aps,preprint]{revtex}
\tighten
\begin{document}
\draft
\title{Quasi-particle spectrum around a single vortex 
in $s$-wave superconductors}
\author{Masaru Kato$^{1,2}$ and Kazumi Maki$^1$} 
\address{$^1$Department of Physics and Astronomy, 
University of Southern California\\
Los Angeles, CA90089-0484\\
$^2$Department of Mathematical Sciences, Osaka Prefecture University\\
1-1 Gakuencho, Sakai, Osaka 599-8531, Japan}
\date{\today}
\maketitle
\begin{abstract}
Making use of the Bogoliubov-de Gennes equation, 
we study the quasi-particle spectrum and the vortex core structure
of a single vortex in quasi 2D $s$-wave superconductors for small $p_F\xi_0$
, where $p_F$ is the Fermi momentum and 
$\xi_0=v_F/\Delta_0$ is the coherence length($\hbar=1$).
In particular we find that the number of bound states decreases rapidly for 
decreasing $p_F\xi_0$. Also for $p_F\xi_0\sim 1$, 
the Kramer-Pesch effect stops around $T/T_c\simeq 0.3$.
\end{abstract}

\section{Introduction}
There are renewed interest in the vortex structure 
since the discovery of the high $T_c$ cuprate superconductors.
Perhaps the superconductivity of high $T_c$ cuprates is characterized 
as $d$-wave superconductivity\cite{tsuei,harlinger} 
and is very close to the quantum limit\cite{maki,doett}.

Schopohl and Maki have studied earlier the quasi-particle spectrum 
around a single vortex line in terms of quasi-classical equation
\cite{eilen,larkin} and predicted a clear four fold symmetry 
for a $d$-wave superconductor\cite{schopohl}. 
Then a beautiful STM study of vortices in YBCO monocrystals was reported
\cite{maggio}.
A big surprise is first of all the quasi-particle spectrum exhibits a circular
symmetry. There is no trace of four-fold symmetry. 
Second, there appears to be only a single bound state 
with energy $\approx \frac{1}{4}\Delta(0)$ where $\Delta(0)$ ($=260{\text K}$) 
is the superconducting order parameter at $T=0{\text K}$. 
A similar bound state energy is observed earlier by a far-infrared 
magneto-transmission from YBCO film by Karra\"{\i} {\it et al.}\cite{karrai}.
If you recall the analysis of bound states around the vortex core by Caroli
and co-workers\cite{caroli}, 
there should be thousands of bound states.
Of course in usual $s$-wave superconductors 
we have $p_F\xi_0\approx 10^3\sim10^4$, where $p_F$ is the Fermi momentum and 
$\xi_0=v_F/\Delta(0)$ is the coherence length($\hbar=1$).
So perhaps the single bound state in YBCO suggests\cite{maki} 
$p_F\xi_0\approx 1$.
At the first sight this suggestion looked outrageous, since this implies
$E_F\approx 200\sim 500{\text K}$ in YBCO.
By analyzing the spin gap at $T=0$K observed in YBCO monocrystals by
inelastic neutron scattering by Rossat-Mignod {\it et al.}\cite{rossat},
we can deduce the chemical potential $\mu$ given as\cite{won},
\begin{equation}
\mu = -345(x-0.45) {\text K}
\end{equation}
where $x$ is the oxygen dopage corresponding to YBa$_2$Cu$_3$O$_{6+x}$.
Then for optimally doped YBCO we obtain $\mu =-190K$.

In D\"{o}ttinger {\it et al.}\cite{doett} the flux flow resistance of 60 K
YBCO measured by Matsuda {\it et al.}\cite{matsuda} is analyzed 
and they identified the Kramer-Pesch effect characteristic to the 
superconductor in the clean limit\cite{K-P,L-O}.
On the other hand apparently the Kramer-Pesch effect is absent 
in 90 K YBCO\cite{matsuda}, implying again perhaps 90 K YBCO
is in the quantum limit\cite{doett}.
Here Kramer-Pesch effect means that the vortex core shrinks 
with decreasing temperature due to the decrease in the occupied 
bound states around the vortex line.
The core size is expressed as\cite{K-P}
\begin{equation}
\xi_1=\frac{v_F}{\Delta(T)}\frac{T}{T_c}.
\end{equation}
This reduction of the core size results to the nonlinear conductivity
\cite{doett,L-O}. Ichioka {\it et al.}\cite{ichi} found the Kramer-Pesch effect
in $d$-wave superconductivity with the help of the semi-classical approach 
developed in Ref.\ \onlinecite{schopohl}.

But if $p_F\xi_0\approx 1$, this implies the semi-classical 
approach introduced in Ref.\ \onlinecite{eilen} and Ref.\ \onlinecite{larkin}
is no longer reliable in studying the vortex in the high $T_c$ cuprates. 
So Morita, Kohmoto and Maki\cite{morita} studied 
Bogoliubov-de Gennes equation for a single vortex in $d$-wave superconductors.
Indeed choosing $p_F\xi_0=1.33$, they can describe
gross features of STM result for YBCO.
On the other hand they have not attempted the 
self-consistency due to the numerical difficulties.
Nevertheless they discovered there is 
a single bound state for $p_F\xi_0=1.33$ 
in the vortex of d-wave superconductors.
Further there are low energy ($E\leq0.1\Delta$) extended states 
with four legs stretched in the four diagonal directions $(1,1,0)$ and 
$(1,-1,0)$.
Recently these extended states are rediscovered by Franz and Te\u{s}anovi\'{c}
\cite{franz} in a model somewhat different  from 
the one in Ref.\ \onlinecite{morita}.
Also in the model of Franz and Te\u{s}anovi\'{c} 
they don't have any bound state.
But since they introduced a large short range Coulomb repulsion,
it is very likely the bound state is completely knocked off.

In order to avoid the mathematical complication related 
to $d$-wave superconductors and to prepare for such a study in $d$-wave
superconductors, we study in this paper the quantum limit 
of a single vortex in $s$-wave superconductors.
Contrary to $d$-wave superconductors the angular momentum  around the vortex
is a good quantum number, which simplifies the problem.
Indeed we can follow the formalism set up in 
Gygi and Schl\"{u}ter\cite{gygi-schl}
except we are now interested in the region of small $p_F\xi_0$.
Recently a related study by Hayashi {\it et al.}\cite{hayasi} is reported.
Unfortunately they have used the $p_F\xi_0$ dependent cutoff energy 
rather than the constant cutoff energy.
So the comparison of systems with different $p_F\xi_0$ appears to have 
involved some artifact. 
One of our purposes is to find the result free from artifact.

In the following we  limit ourselves to $p_F\xi_0=1,2$ and $4$ 
and study the quasi-particle spectrum and 
the shape of $|\Delta(r)|$ as a function of temperature.
Even for $p_F\xi_0=1$ we find  3 bound states at $T=0$K.
Also the Kramer-Pesch effect which is present at $T/T_c\simeq0.5$
appears to stop at $T/T_c=0.3$ for 
$p_F\xi_0=1$.
Finally the shape of $|\Delta(r)|$ appears to depend crucially on the 
$u_0(r)v_0^*(r)$, where $u_0(r)$ and $v_0(r)$  are the wave function 
associated with the lowest bound state energy.
We also show the local quasi particle density of states which 
exhibits remarkable structures.

\section{Bogoliubov-de Gennes equation}
The spatial dependence of order parameters of the superconductivity 
is described by the Bogoliubov-de Gennes equation.
We consider the case with low magnetic field near $H_{c1}$ so that
there is a single vortex. 
Therefore we can ignore the vector potential.
In this case the Bogoliubov-de Gennes equation becomes
\begin{mathletters}\label{BGORIG}
\begin{eqnarray}
\left(-\frac{1}{2m_e}\nabla^2-\mu \right)u_n\left(\bbox{r}\right)+
\Delta\left(\bbox{r}\right)v_n\left(\bbox{r}\right)
=E_nu_n\left(\bbox{r}\right), \\
-\left(-\frac{1}{2m_e}\nabla^2-\mu \right)v_n\left(\bbox{r}\right)+
\Delta^*\left(\bbox{r}\right) u_n\left(\bbox{r}\right)
=E_n v_n\left(\bbox{r}\right),
\end{eqnarray}
\end{mathletters}
where $u_n\left(\bbox{r}\right)$ and $v_n\left(\bbox{r}\right)$ are
quasi-particle wave functions.

We take $z$-axis along to the vortex line.
We consider nearly two-dimensional case for simplicity,
where  kinetic term associated to the  z direction is negligible.
In the following, we merely consider two-dimensional case and 
 we take a cylindrical  coordinate. 
Taking the gauge as $\Delta(\bbox{r})=|\Delta(r)|e^{-i\theta}$ , 
angular momentum of each 
eigenstate becomes half odd integer $m+\frac{1}{2}$\cite{caroli}.
Then $u_n(\bbox{r})$ and $v_n(\bbox{r})$ becomes as follows,
\begin{mathletters}
\begin{eqnarray}
u_n(r,\theta)=u_{n m}(r)\frac{e^{im\theta}}{\sqrt{2\pi}},\\
v_n(r,\theta)=v_{n m}(r)\frac{e^{i(m+1)\theta}}{\sqrt{2\pi}}.
\end{eqnarray}
\end{mathletters}
Following Gygi and Schl\"{u}ter\cite{gygi-schl}, 
we apply Fourier-Bessel expansion with basis;
\begin{equation}
\phi_{mj}(r)=\frac{\sqrt{2} }{RJ_{m+1}(\alpha_{jm})}
J_m(\alpha_{jm}\frac{r}{R})
\end{equation}
to $u_{n m}(r)$ and $v_{n m}(r)$, where $\alpha_{jm}$ is $j$-th positive zero of 
the Bessel function of  $m$-th order  $J_m(x)$.
Then wave functions become as,
\begin{mathletters}
\begin{eqnarray}
u_{n m}(r)=\sum_ju_{n m j} \phi_{m j}(r),\\
v_{n m}(r)=\sum_jv_{n m j} \phi_{m+1 j}(r).\\
\end{eqnarray}
\end{mathletters}
Here, the boundary condition is such that  wave functions are zero 
at the edge of the disk with radius $R$.
Then the Bogoliubov-de Gennes equation becomes,
\begin{mathletters}
\begin{eqnarray}
\left[\frac{1}{2m_e}\left(\frac{\alpha_{jm}}{R}\right)^2
-\mu\right]u_{n m j}+\sum_{j_1}\Delta_{jj_1}v_{n m j_1}
=E_{n m}u_{n m j},\\
-\left[\frac{1}{2m_e}\left(\frac{\alpha_{j\ m+1}}{R}\right)^2
-\mu\right]v_{n m j}+\sum_{j_1}\Delta_{j_1j}u_{n m j_1}
=E_{n m}v_{n m j},
\end{eqnarray}
\end{mathletters}
where $\Delta_{j_1j_2}$ is given as,
\begin{equation}
\Delta_{j_1j_2}=\int_0^R\phi_{m j_1}(r)\left|\Delta(r)\right|
\phi_{m+1j_2}(r)r dr.
\end{equation}
The order parameter is given as,
\begin{equation}
\left|\Delta(r)\right|=
g\sum_{\left|E_n\right|\leq E_c}\sum_{m\ge0}\sum_{j_1j_2}
u_{nm j_1}v_{nm j_2}\phi_{m j_1}(r)
\phi_{m+1j_2}(r)[1-2f(E_{nm})],
\end{equation}
where $g$ is the interaction constant, $E_c$ is the cutoff energy and $f(E)$ is
the Fermi distribution function.
We solve these equations self-consistently.
We fix the cutoff energy ($E_c=5 \Delta_0$ 
for the case of largest Fermi momentum)
and the interaction constant, and choose the 
Fermi momentum $p_F$ so that $p_F\xi_0=$  1, 2 and 4.

\section{Results}
In the numerical calculation, the  radius of boundary is taken as $R=10\xi_0$ 
and maximum angular momentum and number of zero points of Bessel functions are 
taken so that all of the quasi-particle states within cut-off energy  
are taken into the calculation.

\subsection{temperature dependence of order parameter}
\label{sec:op}
The temperature dependence of the order parameter is shown in Fig.\ref{OP}.
Near the boundary $r/\xi_0=10$, 
the order parameter goes to zero and there is a peak before that. 
These are effects from the boundary condition  and the finite size.
But core structures are not affected by the boundary condition.

For $p_F\xi_0=4$ at low temperature ($T/T_c\leq0.1$)
 there is a shoulder in $|\Delta(r)|$ at the vortex core.
This feature comes from bound states in the vortex core.
This can be seen from Fig.\ref{OPsep}, 
where  contributions of the scattering states and the bound states are shown 
separately.
For  $p_F\xi_0=1$, the peak position of 
the contribution to the order parameter is located slightly outside 
of vortex core.
Therefore the core structure is not so much affected by  the bound states.
For larger $p_F\xi_0(\sim 16)$, Hayashi {\it et al.}\cite{hayasi}
 showed the oscillation 
of the order parameter at the inside of the vortex core.
It appears that the  origin of this oscillation is 
the reflection of the bound state wave functions 
contrary to their interpretation.

Slightly increasing temperature, the bound state contribution decreases rapidly
and scattering state contribution remains almost the same. 
In this temperature range, the core size is dominated by the bound states.
Above this temperature region, scattering state contribution decreases with
increasing temperature as the behavior of the order parameter of 
the uniform solution.

We also plot the quasi particle wave function  $u_{n\mu}(r)$ and $v_{n\mu}(r)$
of three  bound states for $p_F\xi_0=4$ in Fig.\ref{amp}.
From these figures, it can be seen that
main contribution to the core structure comes from the lowest energy 
bound state.
Also we can see $u(r)$ of the lowest energy bound state 
behaves like $s$-wave function and $v(r)$ behaves like $p$-wave function.
Similar behavior also can be seen for second and third lowest bound states.
They belong to the angular momentum $m$ and $m+1$ state respectively.

\subsection{Quasi-particle spectrum}
In Fig.\ref{excit} we give  quasi-particle energies with 
angular momentum $m+\frac{1}{2}$ and we also plot those of the uniform state
in order to compare the vortex state and the uniform state.
For small angular momentum, lowest energy states become bound states
lowering its energy and slightly increasing energies of  higher energy states.
For larger angular momentum, the quasi-particle energy is almost same 
for vortex and uniform states and there is no bound state.
Although it is difficult to say the boundary of these two cases 
in the angular momentum, the number of bound states may be counted 
as $20$ for $p_F\xi_0=4$, 7 for $p_F\xi_0=2$ and 3 for $p_F\xi_0=1$.
The temperature dependence of quasi-particle energy 
is shown in Fig.\ref{excit-ft}.
In this figure we plot quasi-particle energies of the several bound states
normalized with the order parameter $\Delta(T)$ at $r=5.5\xi_0$.
From this figure we can see that for smaller $p_F\xi_0$ and 
higher energy states, the quasi-particle energy varies parallel to 
the order parameter with temperature. 
But lower energy states for larger $p_F\xi_0$, the quasi-particle energy 
increases rapidly with decreasing temperature and becomes constant 
at low temperature.
This behavior of the energy of the lowest bound state
was already noted by Gygi and Schl\"{u}ter\cite{gygi-schl} for much larger
$p_F\xi_o$ but the energy increases with decreasing temperature 
even  at low temperature because of the large $p_F\xi_0$.

\subsection{Core radius}
As mentioned in \ref{sec:op}, the core structure depend on the bound states.
As defined by Kramer and Pesch\cite{K-P}, we calculate core radius $\xi_1$ as
\begin{mathletters}
\begin{eqnarray}
\frac{1}{\xi_1}&=&\lim_{r\rightarrow 0}\frac{\Delta(r)}{r\Delta_0},\\
     &=& \frac{g}{\Delta_0}\sum_{E_{nm}\leq E_c}\sum_{j_1j_2}u_{n0j_1}
v_{n0j_2}
\left[1-2f\left(E_{n0}\right)\right]\phi_{j_10}(0)
\frac{d\phi_{j_21}}{dr}(0).
\end{eqnarray}
\end{mathletters}
We plot total $\xi_1^{-1}$ and 
a contribution from scattering states to $\xi_1^{-1}$ in Fig.\ref{xi}. 
The temperature dependence of $\xi_1$ is almost linear in the intermediate
temperature region according to Kramer and Pesch\cite{K-P}.
For $p_F\xi_0=4$, there is a substantial decrease of the core radius
at low temperature because of empty bound states.
Although for smaller $p_F\xi_0$ there is a bound state contribution,
its peak position of the contribution to the order parameter is nearly at the 
edge of the core. 
Therefore the core radius is not much affected.
But for large $p_F\xi_0$, the peak of the contribution of the bound states
are well inside of the core, so the core radius shrinks significantly.

Comparing with Fig.5 in Ref.\ \onlinecite{gygi-schl},
our result for small $p_F\xi_0$ shows saturations of $\xi_1^{-1}$
at $T/T_c\approx 0.1$ for $p_F\xi_0=4$,  
$T/T_c\approx 0.2$ for $p_F\xi_0=2$ and $T/T_c\approx 0.3$ for $p_F\xi_0=1$.
This comes from finiteness of the bound states.

\subsection{Current density and magnetic field}
Current density is calculated  from
\begin{equation}
\bbox{j}(\bbox{r})=\frac{e}{2m_ei}\sum_{nm}\left\{f(E_{nm})
u_{nm}^*(\bbox{r})\nabla u_{nm}(\bbox{r})+\left[1-f(E_{nm})\right]
v_{nm}(\bbox{r})\nabla v_{nm}^*(\bbox{r}) - {\text h.c.}\right\}.
\end{equation}
There is a rotational symmetry, so the current has only $\theta$ component;

\begin{equation}
j_{\theta}(r)=\frac{e}{m_e}\sum_{nm}\left\{f(E_{nm})
\frac{m}{r}\left|u_{nm}(r)\right|^2-
\left[1-f(E_{nm})\right]\frac{m+1}{r}
\left|v_{nm}(r)\right|^2\right\}.
\end{equation}
We show  this current density at $T=0.1T_c$ in Fig.\ref{current} 
for each $p_F\xi_0$.
The peak position of the current density is same as the peak position of 
the lowest energy bound state.
Therefore smaller $p_F\xi_0$ the peak is located further from the vortex core.
Therefore there is no simple relation between the core size and 
the peak position of the current.

From Maxwell equation, we calculate magnetic field parallel to the $z$-axis
as,
\begin{equation}
H_z(r)=\frac{4\pi}{c}\int_r^R j_\theta(r)dr,
\end{equation}
and  we normalized it with $\phi_0/2\pi\lambda^2$ and show in Fig.\ref{mag},
where $\lambda=\left(m_ec^2/4\pi ne^2\right)^{1/2}$ is the penetration depth
and $\phi_0=hc/2e$ is the flux quantum.
From this figure, we find that for smaller $p_F\xi_0$ the distribution of 
magnetic field is more extended to outside of the vortex core.

\subsection{Local density of states}
In order to compare  with the  STM experiments, 
we calculate the local density of states
with  thermal average,
\begin{equation}
N(r,E)=\sum_{nm}\{|u_{nm}(r)|^2f'(E_{nm}-E)+
|v_{nm}(r)|^2f'(E_{nm}+E))\}.
\end{equation}
We also take spatial average with a Gaussian distribution 
with standard deviation $0.1\xi_0$.
This is shown in Fig.\ref{ldos}.
In this figure there is a oscillation at $E/\Delta_0>1.0$, where
$\Delta_0$ is the average value of the order parameter 
at the  outside of the vortex core.
This feature comes from the finite size of our system and it is an artifact.

In addition to this, there are several peaks due to the discrete bound states 
for energy less than the energy gap ($E<\Delta_0$)
and there is a  particle-hole 
asymmetry in contrast to the  result of Gygi and Schl\"{u}ter\cite{gygi-schl}.
Wang and MacDonald\cite{wang} showed this asymmetry by solving lattice
model and also Morita {\it et al.}\cite{morita} and Hayasi 
{\it et al.}\cite{hayasi} pointed out it previously.
For smaller $p_F\xi_0$, this discrete structure is more apparent.

\section{Conclusion}
In summary, we have solved the Bogoliubov-de Gennes equation 
for a single vortex state in s-wave superconductors for small $p_F\xi_0$
and obtain the electronic structure around the vortex core.
We show the effect of the discrete bound states 
to the order parameter structure
, the local density of states, the current density and the magnetic field.
For smaller $p_F\xi_0$, the contribution of the bound state decreases and 
depends on the peak positions of wave functions of 
quasi particles. 
Also the discreteness of the bound states is more apparent for smaller
$p_F\xi_0$.

Recently Sonier {\it et al.}\cite{sonier} claimed that they obtained 
the core size of the vortex in NbSe$_2$ by measuring the peak position of the 
current density.
But there are a few problems in their procedure. 
First, as we have shown there is no simple relation between the core size
and the peak in the current density.
Second even if such a relation exists the relation between these quantities 
are different in the clean limit from the dirty limit result shown in Ref.\ 
\onlinecite{sonier}.
In particular the core size in the clean limit is much smaller than the one
in the dirty limit with respect to the peak position in the current density.
At least they should have analyzed their data in the light of 
the clean limit calculation by Gygi and Schl\"{u}ter\cite{gygi-schl}.

Our calculation has been done on the $s$-wave superconductivity 
in the quantum limit.
For high $T_c$ cuprates, $d$-wave superconductivity is required.
Also in the quantum limit, 
the effects from the constraint of the number conservation
may be large as pointed by van der Marel\cite{marel}, although 
we have fixed $p_F\xi_0$ for all temperature region.
Therefore, the anisotropy of the superconductivity and the constraint of 
the number conservation must be taken into account 
in the study of the vortex in high $T_c$ cuprates.
These are left as future problems.

\acknowledgments
The present  work is supported by NSF under grant number DMR 9531720.
One of us (MK) is supported by the Ministry of Education, Science and Culture
of Japan through  the program of exchange researchers.
He thanks to the hospitality of Department of Physics and Astronomy of 
University of Southern California.
Also MK appreciates useful discussion  with A. Goto and A. Nakanishi.


%
%
 \begin{figure}
 \caption{The temperature dependence of the order parameter $\Delta(r)$
    for $p_F\xi_0=4$ (a), $p_F\xi_0=2$ (b) and $p_F\xi_0=1$ (c).
 The order parameter is normalized with $\Delta_0=\Delta(r=5.5\xi_0, T=0)$
 for each $p_F\xi_0$.}
 \label{OP}
 \end{figure}
 \begin{figure}
 \caption{Contributions from the bound states and the scattering states
to the order parameter for $p_F\xi_0=4$ (a), $p_F\xi_0=2$ (b) 
and $p_F\xi_0=1$ (c), where $\Delta_0=\Delta(r=5.5\xi_0, T=0)$.}
 \label{OPsep}
 \end{figure}
 \begin{figure}
 \caption{The quasi particle wave functions $u(r)$ and $v(r)$ 
and their product 
with thermal factor for the first (a), the second (b) and the third (c) 
of the lowest energy bound states.}
 \label{amp}
 \end{figure}
 \begin{figure}
 \caption{The quasi particle spectrum of the vortex state($\Diamond$)  
and the uniform state(dashed line)
for $p_F\xi_0=4$ (a), $p_F\xi_0=2$ (b) and $p_F\xi_0=1$ (c).}
 \label{excit}
 \end{figure}
 \begin{figure}
 \caption{The temperature dependence of the quasi-particle energies 
of the bound states in the vortex state.
The energies are normalized by the order parameter $\Delta(T)$ 
at $r/\xi_0=5.5$. (a), (b) and (c) are for $p_F\xi_0=4$, $p_F\xi_0=2$
and  $p_F\xi_0=1$, respectively}
 \label{excit-ft}
 \end{figure}
 \begin{figure}
 \caption{The temperature dependence of 
$\xi_1^{-1}$ for $p_F\xi_0=1,2$ and $4$.
Also contributions from scattering states for each $p_F\xi_0$ are plotted.}
 \label{xi}
 \end{figure}
 \begin{figure}
 \caption{The current density at $T=0.1T_c$ for $p_F\xi_0=4$ (a), 
 $p_F\xi_0=2$ (b) and $p_F\xi_0=1$ (c). The current density is normalized with 
its maximum value. The order parameter is also shown for comparison.}
 \label{current}
 \end{figure}
 \begin{figure}
 \caption{ The spatial dependence of the magnetic field $H$ at $T=0.1T_c$
  for $p_F\xi_0=1,2$ and $4$. }
 \label{mag}
 \end{figure}
 \begin{figure}
 \caption{The local density of states $N(r,E)$ at $T/T_c=0.1$ for 
 $p_F\xi_0=4$ (a), $p_F\xi_0=2$ (b) and $p_F\xi_0=1$ (c). 
 $N(r,E)$ is normalized with the density of states of the normal state $N_0$.}
 \label{ldos}
 \end{figure}

%
%

\end{document}